\documentclass[useAMS,usenatbib]{mn2e}

\usepackage{graphicx,amssymb,color,amsmath}
\usepackage[normalem]{ulem}

\title[Binary influence on post-MS multi-planets]
{Binary star influence on post-main-sequence multi-planet stability}

\author[Veras, Georgakarakos, Dobbs-Dixon \& G\"{a}nsicke]{
Dimitri Veras$^{1}$\thanks{E-mail: d.veras@warwick.ac.uk},
Nikolaos Georgakarakos$^{2}$,
Ian Dobbs-Dixon$^{2}$,
Boris T. G\"{a}nsicke$^{1}$
\\
$^{1}$Department of Physics, University of Warwick, Coventry CV4 7AL, UK
\\
$^{2}$New York University Abu Dhabi, Saadiyat Island, P.O. Box 129188, Abu Dhabi, UAE
}

\pubyear{2016}

\begin{document}
\label{firstpage}
\pagerange{\pageref{firstpage}--\pageref{lastpage}}
\maketitle

\begin{abstract}
Nearly every star known to host planets will become a white dwarf, and nearly 100
planet-hosts are now known to be accompanied by binary stellar companions. Here, we
determine how a binary companion
triggers instability in otherwise unconditionally stable single-star two-planet systems during 
the giant branch and white dwarf phases of the planet host. We perform about 
700 full-lifetime (14 Gyr) 
simulations with A0 and F0 primary stars and secondary K2 companions, and identify 
the critical binary distance within which instability is triggered at any point
during stellar evolution. We estimate this distance to be about seven times the outer
planet separation, for circular binaries. Our
results help characterize the fates of planetary systems, and 
in particular which ones might yield architectures that are conducive to generating
observable metal pollution in white dwarf atmospheres.
\end{abstract}

\begin{keywords}
minor planets, asteroids: general -- stars: white dwarfs -- methods:numerical -- 
celestial mechanics -- planet and satellites: dynamical evolution and stability
\end{keywords}

\section{Introduction}

Single white dwarf planetary systems provide unparallelled insights into planet formation and
long-term dynamics. Rocky debris is found in the atmospheres of between one-quarter and
one-half of all known single Milky Way white dwarfs \citep{zucetal2003,zucetal2010,koeetal2014}.
The chemical composition of this debris can be compared directly to solar system asteroid
families, and primarily reflect the inner solar system \citep{gaeetal2012,juryou2014,xuetal2014,wiletal2016}.
Detected around a few per cent of these ``polluted'' white dwarfs are dusty circumstellar discs
\citep{faretal2009,beretal2014,baretal2016,farihi2016} -- some of which contain gaseous components
\citep{gaeetal2006,gaeetal2008,meletal2012,wiletal2014,manetal2016} --  and one white dwarf (WD 1145+017) features at least 
one actively disintegrating minor planet 
\citep{vanetal2015,aloetal2016,gaeetal2016,garetal2016,guretal2016,rapetal2016,redetal2016,xuetal2016,veretal2016b,zhoetal2016}.

In a system-wide case of pinball \citep{bonwya2012},
planets are thought to represent a necessary perturbing presence for asteroids, moons or
rocky debris to be thrust towards the white dwarf \citep{bonetal2011,debetal2012,frehan2014,antver2016,
payetal2016a,payetal2016b,veretal2016a}. As objects encounter the white dwarf Roche, or disruption 
radius, they break up and form a disc \citep{graetal1990,jura2003,debetal2012,beasok2013,
veretal2014a,veretal2015a,veretal2016b} and eventually accrete onto the white dwarf 
\citep{bocraf2011,rafikov2011a,rafikov2011b,metetal2012,rafgar2012}. Although the planets
themselves rarely collide with the white dwarf
\citep{veretal2013a,musetal2014,vergae2015,veretal2016a,veras2016b}, 
they are necessary agents to transport smaller debris towards the star.

The above behaviour need not be restricted to single stars. Circumstellar metal pollution, debris discs and 
disintegrating asteroids should all occur in some fraction of binary star planetary systems as well \citep{zuckerman2014}. 
Planets themselves may be common in binary star systems. Above 140 exoplanets are known to reside in these
systems, and about four-fifths of those feature a single planet-host star\footnote{http://exoplanets.org}.
Although a significant body of work has attempted to understand the formation and evolution of these exoplanets, 
little attention has been given to their fates \citep{kraper2012,bonver2015,hampor2016,petmun2016}.
Further, despite the more common case of exoplanets in ``circumstellar'' configurations (orbiting one star),
the rarer ``circumbinary'' case (e.g. \citealt*{doyetal2011,armetal2014,maretal2014,geoegg2015}) has received the 
lion's share of investigations assessing post-main-sequence
fate \citep{sigurdsson1993,vertou2012,musetal2013,portegieszwart2013,schdre2014,voletal2014,kosetal2016}.

The circumstellar case, in addition to being more common than the circumbinary case, can yield more insights.
If binary stars are separated by more than a few tens of au, then they evolve effectively independently
of each other. After one of these stars has become a white dwarf, this separation is large enough to ensure
that the companion's wind provides a negligible chemical change to the white dwarf's atmosphere
\citep{verxu2016}. Consequently, one can link this white dwarf with the characteristics of a single white dwarf planetary system.

Here, we address the dearth of work investigating circumstellar planetary systems within
evolved binaries by estimating a stability limit for a relatively simple but important base case.
We consider
two nearly coplanar planets in each system, as opposed to the single planet cases
from \cite{kraper2012}, \cite{bonver2015} and \cite{hampor2016}, and the high mutual inclinations
between the binary star and planetary system in \cite{petmun2016}.  Our focus is also entirely
different. \cite{kraper2012}
detailed how a planet may ``hop'' from one star to the other during post-main-sequence mass loss,
and \cite{bonver2015} illustrated how the combination of Galactic tides and post-main-sequence mass loss
conspire to stretch the orbits of stellar binaries such that a previously quiescent planet becomes
suddenly dynamically active. Our two planets orbit the same star at distances commensurate
with those of solar system
giant planets, which precludes star hopping. Our initial binary separation is no greater than 500 au, 
rendering Galactic tides ineffective.

Consequently, our focus is on the stellar phase-dependent stability of the resulting four-body system.
We determine simply the binary separation within which the two-planet system becomes unstable during their
host star's main sequence, giant branch and white dwarf phases, rather than introducing more complex considerations such as Lidov-Kozai oscillations or accretion rate estimations \citep{hampor2016,petmun2016}.  In
Section 2 we briefly recapitulate
some relevant stability results, before in Section 3 detailing the initial conditions for our simulations.
We present our results in Section 4 and conclude in Section 5.

\section{Stability limits}

Unlike the three-body problem, for which many analytic formulations of stability exist \cite[e.g.][]{georgakarakos2008},
similar concise expressions for four-body problems are rarer \citep[e.g.][]{lokser1985,serlok1987}, even despite major
recent progress with central configuration theory \citep{erdczi2016,hamilton2016,veras2016c}. Further, when one
of the bodies loses mass -- as is the case for a main-sequence star which becomes a white dwarf -- then
not even the two-body problem is solvable (see Chapter 4 of \citealt*{veras2016a}).

These restrictions persuade us to construct binary systems based on our knowledge of the three-body problem
with mass loss. Two-planet systems where the central star evolves into a white dwarf were the focus of investigations
by \cite{debsig2002}, \cite{veretal2013a} and \cite{voyetal2013}. A key finding from \cite{debsig2002} was that
two planets which are Hill stable on the main sequence -- meaning that their orbits could never cross --
might no longer be Hill stable as the central star loses mass. \cite{veretal2013a} then expanded on that work
with a wider exploration of parameter space, and by also considering Lagrange stability, which is the
dynamical state of two Hill stable planets that remain bounded and ordered. \cite{voyetal2013} explored the
evolution of two planets in mean motion resonances, and the ``non-adiabatic'' orbital variations resulting
from violent mass loss.

The adiabaticity of a system is defined here as the extent to which its planets' orbits change in a 
predictable manner due to stellar mass loss: when the eccentricity variations are negligible and
the semimajor axes values increase in direct proportion to the extent of the mass loss.
Adiabaticity is quantified by a formula like that of equation (15) of \cite{veretal2011},
which represents a scaled ratio of the orbital period to the mass loss timescale. When this ratio
is much less than unity, the system is said to be adiabatic. \cite{veretal2011} showed that any
stellar companion within several hundred au will evolve adiabatically. The stability boundary
changes found by \cite{debsig2002} and \cite{veretal2013a} hence considered only the adiabatic
state because that corresponds to the expected fate of the majority of currently-known
exoplanetary systems.

The Hill stability limit and probably the Lagrange stability limit are scale-free in the sense
that they depend on ratios of orbital elements of the planets rather than the explicit 
planet-star separations themselves \citep{gladman1993,bargre2006,donnison2006,bargre2007,rayetal2009,
vermus2013,marzari2014,petrovich2015}. This property is particularly useful for studies such as this one in order to 
reduce the number of degrees of freedom of the explored systems and to place the planets
well-beyond the tidal reach of the evolving star \citep{villiv2009,kunetal2011,musvil2012,
adablo2013,norspi2013,viletal2014,staetal2016}.  However, after instability does
occur, the problem no longer remains scale-free \citep{johetal2012,petetal2014}.

Of particular interest here are two-planet systems which are sufficiently Lagrange stable
around a single star to survive Gyrs into the white dwarf phase, but not under the influence
of a binary companion. Stellar mass loss will expand the orbits of both planets
relative to their initial values more than the orbit of the binary companion
(even though all expansions are adiabatic; see equation 3 of \citealt*{kraper2012},
equations A22-A23 of \citealt*{petmun2016} and equations 7.6-7.7 of \citealt*{veras2016a}).
Consequently, because angular momentum must be conserved (assuming isotropic mass loss),
the post-mass-loss state of the system leaves it more fragile and susceptible to instability
than on the main sequence.

\section{Initial conditions}

\begin{figure}
\includegraphics[width=8cm]{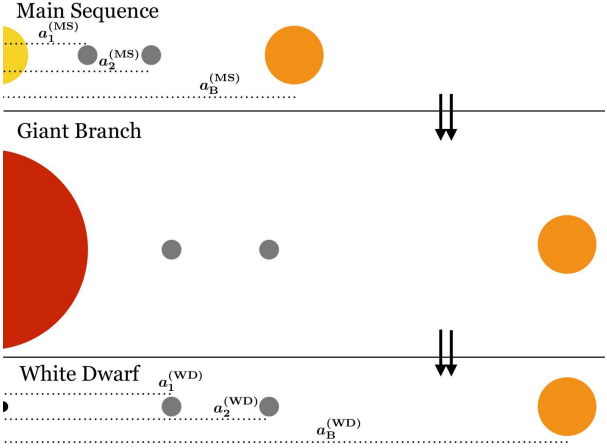}
\caption{
Cartoon (not to scale) of our setup: two planets (small grey circles) orbiting
the primary (yellow, red then black star at left), plus a secondary 
(orange star at right).
As the primary evolves from the main sequence (top region) to the giant branch phase
(middle region) to the white dwarf phase (bottom region), the star expands and then contracts.
Concurrently, the planets and secondary all expand their orbits. This paper explores
what binary separations would create instability, and during what phases.
}
\label{cart}
\end{figure}

The above considerations motivate our initial conditions, which are measured in Jacobi
coordinates. Assume that the primary star is the
planet-hosting star and evolves from a main sequence star to a white dwarf. Denote its mass
as $M_{\star}(t)$. The two planets it hosts have masses $M_1$ an $M_2$ and semimajor axes
$a_1(t)$ and $a_2(t)$ such that $a_1(0)<a_2(0)$. The binary stellar companion, which will be
denoted as the ``secondary'', has mass $M_{\rm B}$
and a semimajor axis of $a_{\rm B}(t)$. The secondary's mass is
time-independent because it is assumed
(i) to be low enough such that it remains on the main sequence for the duration of our simulations,
and (ii) mass loss from main sequence winds is negligible.
In Fig. \ref{cart}, we show a 
cartoon of the bodies in our simulations and how they would evolve in time in a stable manner.

The large phase space of the four-body problem, combined with the time-consuming nature of the
simulations that extend beyond the main sequence, demanded that we carefully chose what parameters
to explore. We focused on $a_{\rm B}(t)$ and minimally sampled variations of other parameters.

\subsection{Masses}

We fixed $M_1 = M_2 = M_{\rm Jupiter}$ and $M_{\rm B} = 0.81M_{\odot}$, which corresponds to a K2 star
by Appendix B of \cite{gray2008}. Using that same table we sampled both A0 stars ($M_{\star} = 2.34M_{\odot}$)
and F0 stars ($M_{\star} = 1.66M_{\odot}$) for our primary. These choices allowed us to evolve the primary
star through all phases of stellar evolution while the secondary remained on the main sequence.
The main sequence lifetimes of A0, F0 and K2 stars are respectively, roughly $0.75$, $2$ and~$>$~$20$~Gyr; 
the A0 and F0 stars turn into white dwarfs with masses of about $0.65M_{\odot}$
and $0.60M_{\odot}$ \citep{huretal2000}. Both A0 and F0 stars represent common progenitors of the present-day
population of white dwarfs in the Milky Way \cite[e.g.][]{treetal2016}.

\subsection{Orbits}

We set $a_1(0) = 5$ au and sampled three values of $a_2(0)$ corresponding
to 8, 10 and 20 times the mutual Hill radius ($\beta$), as defined by \cite{smilis2009}. For A0 stars,
these values correspond to $a_2(0) = 8.50, 9.79, 23.4$ au, whereas for F0 stars, they correspond to 
$a_2(0) = 9.10, 10.7, 31.5$ au. The resulting semimajor axis ratios are far from any strong orbital 
period commensurability, and large enough to ensure Hill
stability \citep[e.g.][]{donnison2006} and probably Lagrange stability \citep{petrovich2015},
{\it both} along the main sequence and white dwarf phases \citep{veretal2013a}. Further, 
because $M_{1}/M_{\star}, M_{2}/M_{\star} \ll 1$ during all phases, the breakdown of these 
limits for sufficiently high planet masses \citep{morkra2016} is not consequential here. 
We then set the planetary initial orbital eccentricities to zero. These
choices help us focus on determining the global stability boundary as a function of $a_{\rm B}$
and not deplete resources by modelling already-unstable two-planet, one-star systems.

We sampled up to eleven different values of $a_{\rm B}$ for a given set of mass and planetary
separations, and adopted the range $50 \ {\rm au} \le a_{\rm B}(0) \le 500 \ {\rm au}$. For stability
considerations, the absolute value of $a_{\rm B}(0)$ is not as important as the ratio
$a_{\rm B}(0)/a_{2}(0)$, through which we report our results. However, the upper bound of $a_{\rm B}(0)$
is important because it is low enough that we can justify ignoring the effect of Galactic
tides and stellar flybys \citep{veretal2014b}. Although our focus was on 
circular binary orbits, we also performed simulations with nonzero binary eccentricities ($e_{\rm B}$) of
0.2, 0.4 and 0.8.

Due to the chaotic nature of long-term evolution simulations of planetary systems,
we did not wish to perform just one simulation for each set of masses, semimajor axes and eccentricities sampled.
Instead we performed four per set. Within each of the four simulations, for the two planets and the binary companion, we randomly selected, from a uniform distribution, mean anomalies, arguments of pericentre and longitudes of ascending node across their entire
ranges, and inclinations between $-1^{\circ}$ and $1^{\circ}$. \footnote{This choice helps prevent an artificially
high rate of planet-planet collisions; see \cite{musetal2014}.} Effectively, both stars and both planets are all nearly coplanar.

\subsection{Integrator}

In order to simulate system evolution across all stellar phases, we needed to concurrently propagate
the orbital evolution of the planets and binary star with the physical evolution of the primary.
A currently available tool which is well-suited for this purpose is the modified version of the 
Bulirsch-Stoer integrator from the {\tt Mercury} integration package \citep{chambers1999}.
The modifications were detailed in \cite{veretal2013a} and involve splicing the mass and radius
output from the {\tt SSE} stellar evolution code \citep{huretal2000} within each 
Burlisch-Stoer substep in order to determine each global timestep. 

The consequence is that tens or hundreds of timesteps are needed to resolve a single orbit of the
inner planet. Long-term integrations are hence time-consuming.
We integrated our simulations for 14 Gyr, which is the current age of the Universe (also known
as a Hubble time), and were able to perform a total of about 700 simulations with two stars
and two planets. We then repeated all simulations but with the binary star removed. Doing so provided
a necessary contrast in order to isolate the binary influence.

Because the primary was treated as a point particle with a variable mass, mass loss is effectively 
treated as isotropic, a well-used and justified approximation \citep{veretal2013b}. The ejection
distance is computed using the Galactic tidal model from \cite{vereva2013} and assuming that
the stars reside at a distance of 8 kpc from the Galactic centre. This ejection boundary takes the form 
of an ellipsoid, and an ejection was deemed to occur when the code detected that a planet resided
outside of this ellipsoid. The ellipsoid itself changes shape as the primary loses mass; this change
was tracked by our code. The primary was evolved
assuming Solar metallicity, a Reimers mass loss numerical coefficient of 
$2 \times 10^{-13} M_{\odot}$~yr$^{-1}$ for the classic prescription \cite{kudrei1978},
and, along the asymptotic giant branch, the superwind prescription of \cite{vaswoo1993}.

\begin{figure}
\includegraphics[width=8cm]{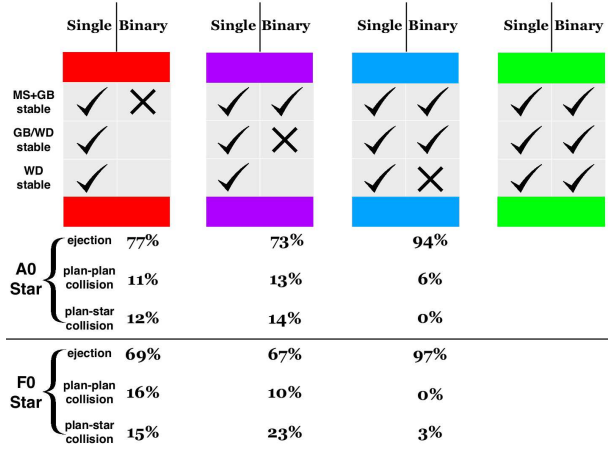}
\caption{
  Classification of our results. The designations MS, GB and
  WD refer to main sequence, giant branch and white dwarf,
  and ``GB/WD stable'' refers to the timespan of 10 Myr
  which bisects the transition point between the two phases.
  The colours refer to particular configurations which feature
  instability: {\it red} for binary system instability on the main
  sequence or giant branch phase, {\it purple} for binary system
  instability during the GB/WD transition, and {\it blue} for
  binary system instability during the white dwarf phase.
  {\it Green} refers to binary systems which remained stable throughout
  14 Gyr of evolution. All single-star planetary systems remained
  stable over this timespan.  The instability statistics
  below the chart are divided according to whether the primary
  is an A0 star or F0 star.
}
\label{tabsum}
\end{figure}

\begin{figure*}
\centerline{
\includegraphics[width=8cm]{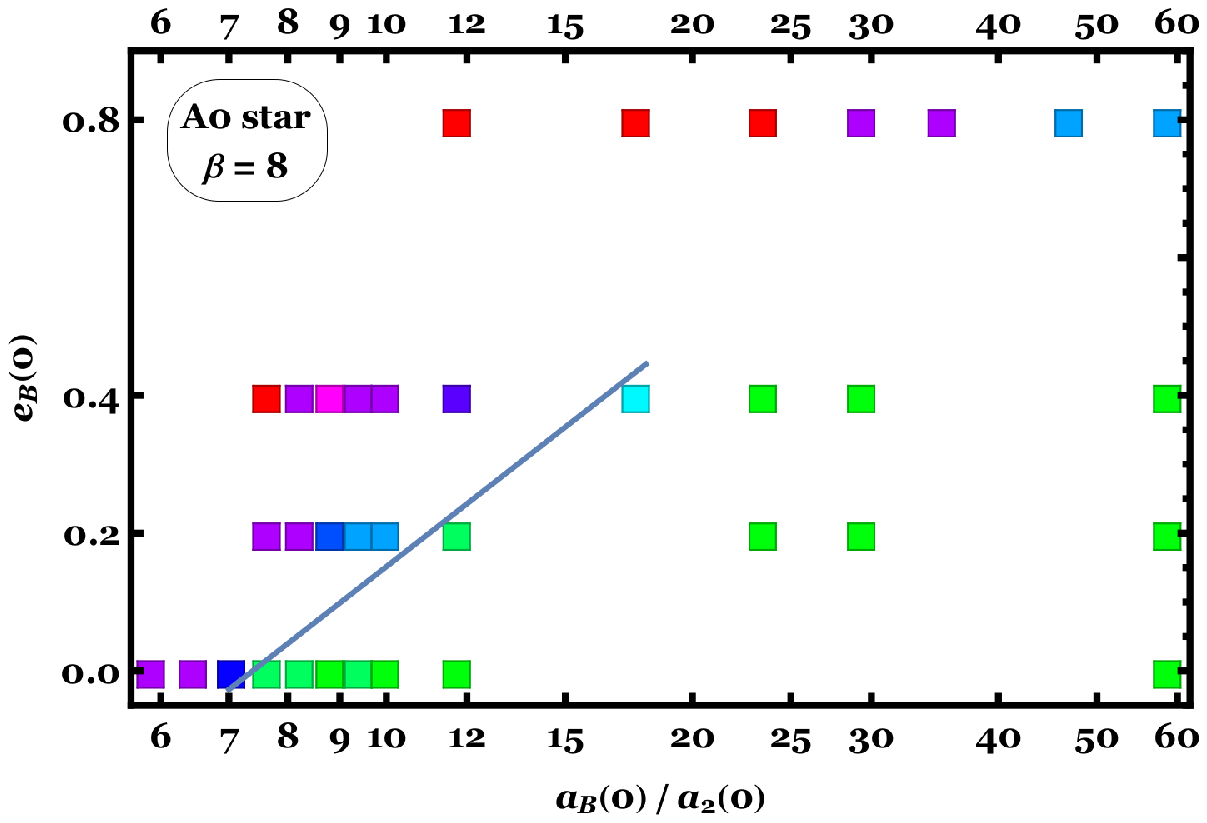}
\includegraphics[width=8cm]{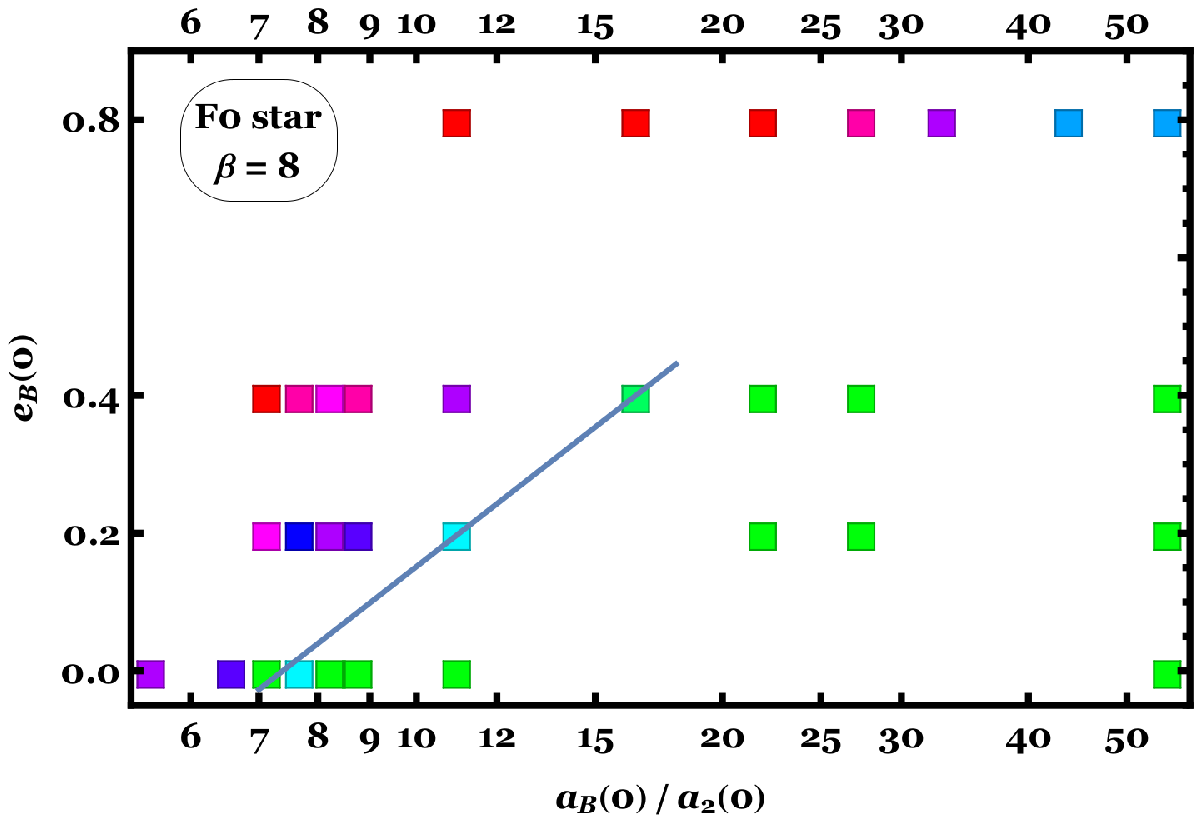}
}
\centerline{
\includegraphics[width=8cm]{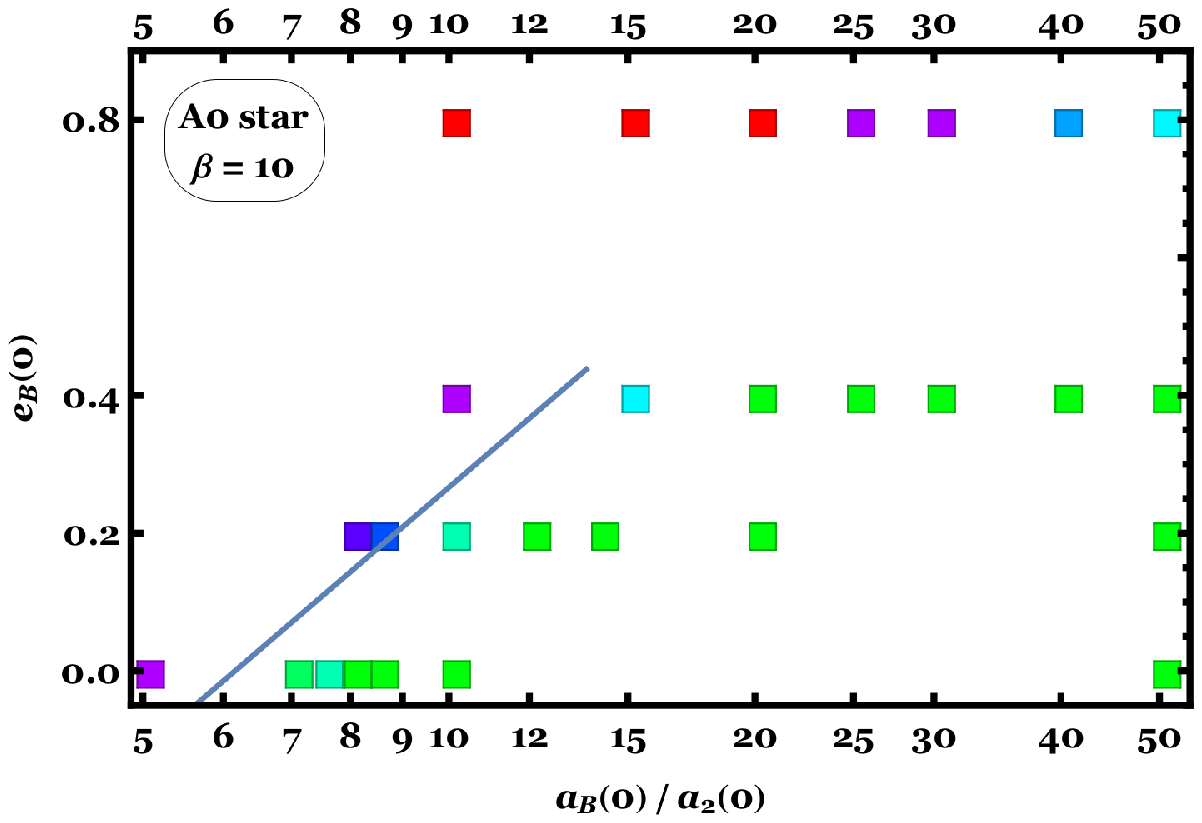}
\includegraphics[width=8cm]{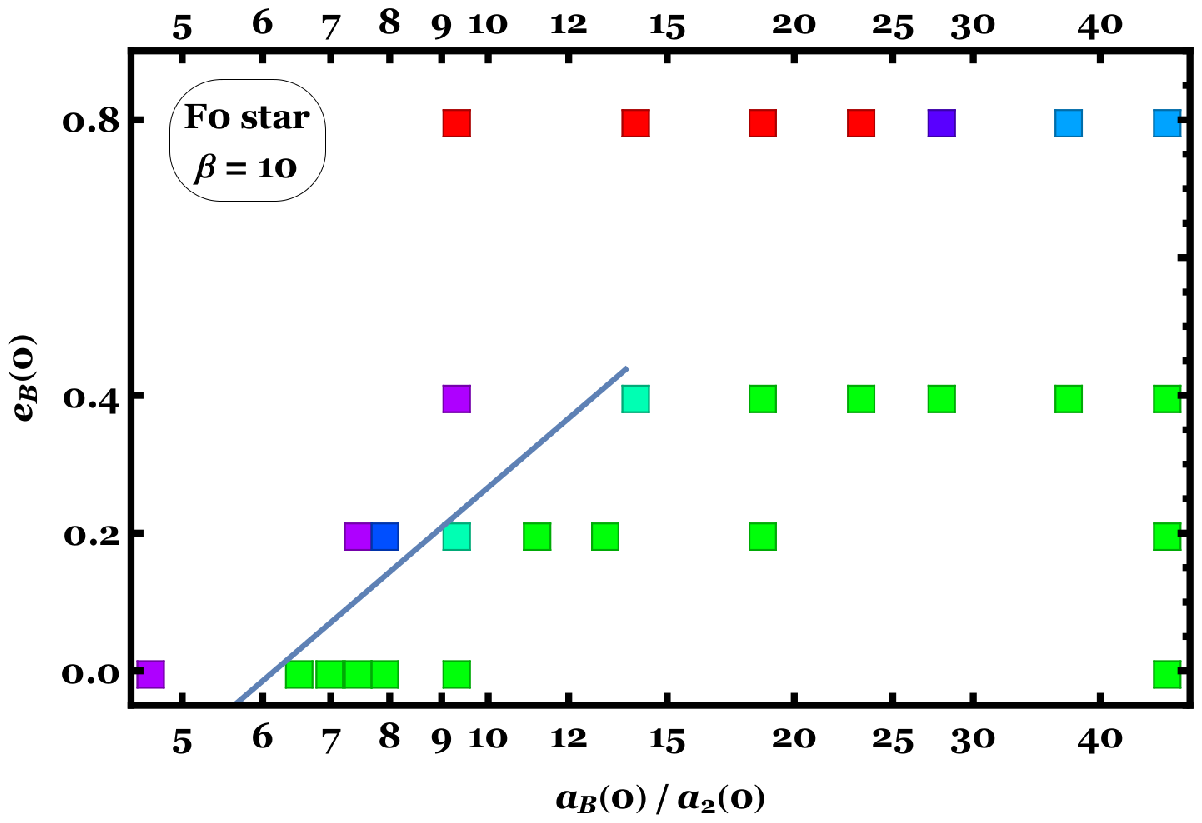}
}
\centerline{
\includegraphics[width=8cm]{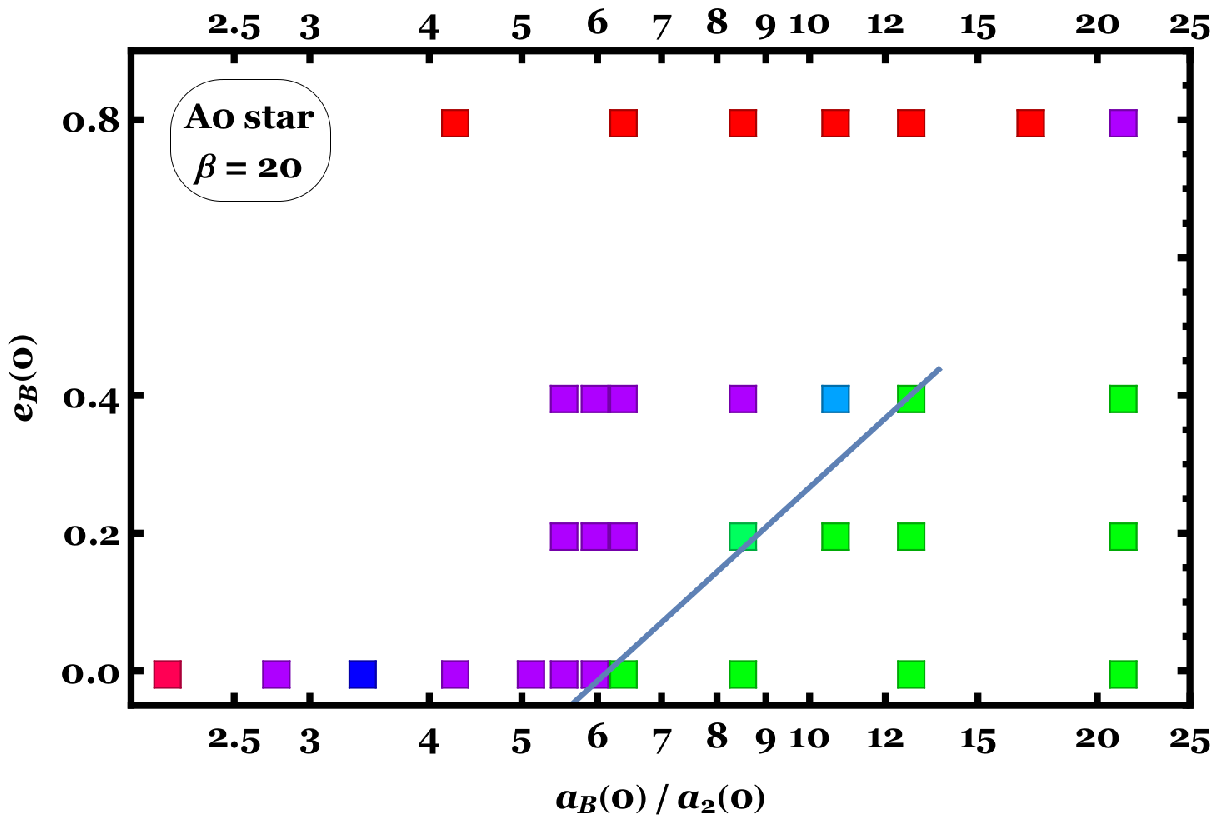}
\includegraphics[width=8cm]{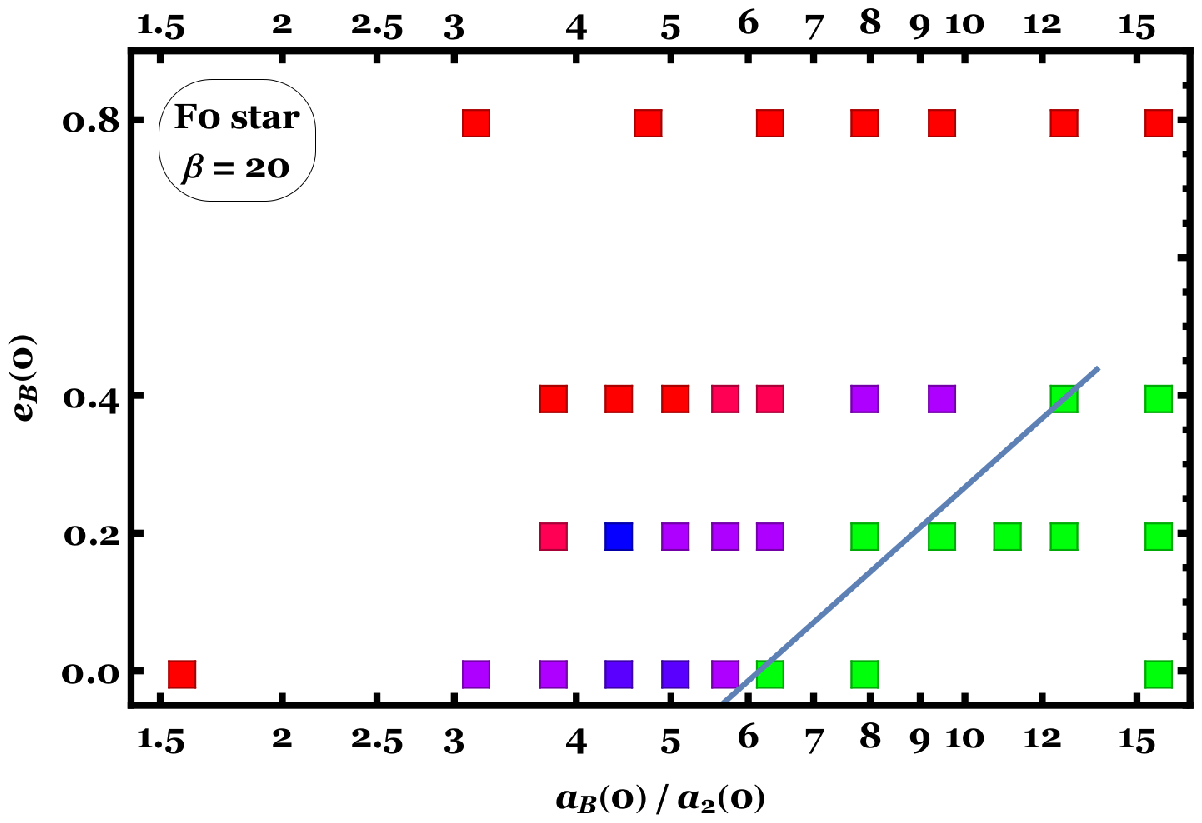}
}
\caption{
  Stability limits ({\it red}: unstable on main sequence or giant branch phases; {\it purple}: unstable
  during the giant branch / white dwarf transition; {\it blue}: unstable on white dwarf phase; {\it green}:   stable throughout). The squares represent colour-blended
  outcomes (see Fig. \ref{tabsum}) of sets of four simulations run with different
  initial orbital angles and small inclinations. The diagonal lines correspond to rough
  empirical estimates (equation \ref{piecewise}) for the transition between green squares
  and the other squares. The left and right panels, which are consistent with one
another, respectively 
contain A0 and F0 primary stars. In all panels, $a_1(0)=5$~au; the top, middle and
bottom panels have, for A0 primary stars, $a_2(0) = 8.50, 9.79, 23.4$~au, and for F0 
primary stars,
$a_2(0) = 9.10, 10.7, 31.5$~au.  These values correspond to mutual Hill
radii separations of $\beta = 8, 10$ and 20. The plots suggest that (i) for circular
binaries, $a_{\rm B}(0)/a_2(0) > 7$ nearly ensures stability, and (ii) for highly eccentric
binaries, the good colour resolution over a wide range of $a_{\rm B}(0)/a_2(0)$
allows one to more easily generate instability at a desired phase for a given
set of initial conditions.
}
\label{mainplot}
\end{figure*}

\section{Results}

Our goal is to link binary separation with stability. We define an unstable system as
one which has experienced an ejection or collision. In our simulations, the collisions
occurred either between two planets or between a planet and the primary (the
planetary radius was equivalent to that of Jupiter and the stellar radii were given
by {\it SSE}). Only planets
were ever ejected: all secondaries remained bound to the primary. Every one of our 
set of simulations with no secondary remained stable for 14 Gyr, providing a useful
point of comparison and confirming that our choices of $a_1(0)$ and $a_2(0)$ were
Lagrange-stable in the single-star case.

\subsection{Instability classification}

We classified simulation outcomes according to four types, which are displayed
in Fig. \ref{tabsum}. Each type is colour-coded for easy linkage to our main result,
which is presented in Fig. \ref{mainplot}. The types are: (green) stable throughout
the simulation; (blue) stable until some point along the white dwarf stage; (purple)
stable until the transition period between the end of the giant branch and start of the
white dwarf phase; (red) unstable during the main sequence or giant branch phases.
We define the transition period as the 10 Myr span which is bisected by the exact
time when the star becomes a white dwarf (964 Myr for A0 stars and 2270 Myr 
for F0 stars, according to \citealt*{huretal2000}). This transition period includes
the tip of the asymptotic giant branch, where the ``superwind'' \citep{lagzij2008} is 
strongest. Consequently, instability is common during this time (e.g. see Appendix A
of \citealt*{veretal2016a}). In contrast, the transition between the main sequence
and giant branch phases is gradual and not dynamically noteworthy.

The instability fractions given in Fig. \ref{tabsum} indicate that the majority
of instability in all cases occurs in the form of ejections. This fraction is above
90 per cent for systems which become unstable during the white dwarf phase. Just a few
per cent of white dwarf unstable systems feature planet-primary collisions,
although this value might be an underestimate because no tidal effects were included
in the simulations, and because collisions were determined according to the actual
white dwarf radius as opposed to its Roche, or disruption, radius. Those
fractions are further slightly blighted by small number statistics (90 total
blue instabilities, as opposed to 148 red and 217 purple ones).

\subsection{Main result}

Our main result, in Fig. \ref{mainplot}, illustrates when instability occurs in binary
separation space. Each square represents the mean outcome of four sets of simulations
run with different mean anomalies, longitudes of pericentre and arguments of ascending node.
The mean is computed by a colour-blend of the four outcomes, where the colours associated
with each outcome are illustrated in Fig. \ref{tabsum}. The bubble in the upper-left corner
of each plot indicates the primary stellar type, and the initial separation of planets
in mutual Hill radii ($\beta$). 

Overall, the figure illustrates a trend from red (main sequence instability) to green
(stability) as the binary separation increases, an expected result. Because the colour trend
is not strictly monotonic (e.g. left side of the upper-rightmost plot) and computational
limitations prevent us from covering the entire phase space, we cannot identify specific
values of $a_{\rm B}(0)/a_{2}(0)$ for which transitions in behaviour occur. Nevertheless,
we can still establish some fairly robust estimates.

Consider first the case for circular binaries. In all cases the systems are stable for
$a_{\rm B}(0)/a_{2}(0) \gtrsim 10$, and probably $a_{\rm B}(0)/a_{2}(0) \gtrsim 7$ given
the bounding $\beta = 8$ and $\beta = 20$ cases. If not stable, then the system 
is likely to survive over the entire main sequence and giant branch phases, but
is unlikely to survive intact well into the white dwarf phase: the transition from giant
branch star into a white dwarf triggers instability unless $a_{\rm B}(0)/a_{2}(0) > 3$.

The same relations cannot be used for more eccentric binary cases. The critical separation
for stability across all phases increases with increasing $e_{\rm B}(0)$. For moderate eccentricity
cases ($e_{\rm B}(0) = 0.2, 0.4$) and $\beta \le 10$, stability is always achieved for
$a_{\rm B}(0)/a_{2}(0) \gtrsim 60$. In contrast, for this semimajor axis ratio
and $e_{\rm B}(0) = 0.8$, instability is achieved predominately along the white dwarf
phase. Highly eccentric binaries appear to stretch out the instability regimes across
a wider swath of $a_{\rm B}(0)/a_2(0)$ space, which fortuitously allows one to 
more easily select initial conditions with predictable outcomes.

In order to aid estimation of the phase space region where the transition 
from stable (green) to unstable (all other colours) occurs for non-zero binary eccentricity,
we provide rough empirical estimates in equation form. We find that stability
is achieved if the following is satisfied:

\[
  e_{\rm B}(0) <
\begin{cases}
  0.50 \ln{\left[\frac{a_{\rm B}(0)}{a_2(0)}\right]} - 1
       & \text{if} \ \beta= 8,      \\[2ex]
  0.55 \ln{\left[\frac{a_{\rm B}(0)}{a_2(0)}\right]} - 1  
       & \text{if} \ \beta= 10, 20
\end{cases}
\]

\begin{equation}
  \label{piecewise}
\end{equation}

\noindent{}for $0.0 < e_{\rm B}(0) \le 0.4$. We have drawn these curves on
all of the plots in Fig. \ref{mainplot}.

We can also compare results for both types of primary star studied (left and right panels).
The agreement is good enough in nearly all cases (but not necessarily the 
$\beta = 20$, $e_{\rm B}(0) = 0.4$ case) to conclude that the stability limits change 
little depending on which type of star in the A-F spectral type range which is used.
This robustness provides evidence that adopting a single primary mass may
be sufficient for future investigations.

\subsection{Eccentricity evolution}

Instability in multi-planet systems, partly through eccentricity generation, may 
easily yield configurations which can
lead to white dwarf pollution, as detailed by 
\cite{veretal2013a}, \cite{musetal2014}, \cite{vergae2015},
\cite{payetal2016a}, \cite{payetal2016b},
\cite{veras2016b}, and \cite{veretal2016a}.
However, what about simulations which remain stable throughout all stellar phases
but are still perturbed due to the presence of a secondary? In the single-star case,
Figure 8 of \cite{musetal2014} illustrates that for their stable simulations, mutual
interactions of three planets
during mass loss can raise eccentricities up to 0.2.
\cite{voyetal2013} used a different integrator and found that eccentricity variations
can easily reach several tenths (see, in particular, their figs. 16-17).

Here, our green-coloured simulations allow us to evaluate how high the planetary 
eccentricities become on the white dwarf phase for stable simulations.  We find
that in extreme cases, planet eccentricities can reach values of about 0.1 from main sequence
maximum values of 0.01. Our more muted
excitation probably arises from the fact that we have adopted much less massive stars than
\cite{voyetal2013} and \cite{musetal2014}, who partly used particularly violent
and rare $8M_{\odot}$ (B2) primary stars.  Further, if our secondary was too far
away to ever trigger instability, then its effect on the planetary eccentricity was not
great enough to raise it to 0.1. Analytical explanations for three-body (and four-body)
dynamics during
stellar mass loss is sorely in need of future exploration, and appears to not represent
a simple extension of two-body dynamics.\footnote{Two-body dynamics, from equations
15 and 17 of \cite{veretal2011}, indicates that the mass-loss induced eccentricity jump
for our simulations would be on the order of just $10^{-4}$. In this situation, there are
no multi-planet interactions.}

Whatever eccentricity excitation occurs during mass loss, after the star has become a white
dwarf, then the system again becomes a fixed-mass system for which other theories may
be applicable. One option is Laplace-Lagrange secular theory (Chapter 7 of \citealt*{murder1999}),
which is sufficient for a first evaluation of the dynamics of a system.  However, this
theory is of low order in the masses, eccentricities and semimajor axis ratio $a_1/a_2$. Hence,
it may give reasonable results for problems in our Solar Sytem but typically not in exoplanetary
systems (where eccentricities may be high and semimajor axis ratios small) or for planets in
stellar binaries. In that latter case, there are several studies that perform expansions to
second order in the masses and develop solutions for higher values in the eccentricities and
in the semimajor axis ratio \citep[e.g.][]{georgakarakos2002,georgakarakos2003,libetal2005,georgakarakos2006,libetal2006,verarm2007,georgakarakos2009,giuetal2011,libsan2013,georgakarakos2016},
which can describe the dynamics of the system more accurately.  Of course,
as the above studies refer to three-body systems, they could only be used if our
four-body system could be approximated by a three-body one (as for example
where the companion star is sufficiently distant to be considered as having a negligible
influence on the orbital evolution of the two planets).

\section{Conclusion}

Motivated by the fact that planetary origins of metal pollution in white dwarf 
atmospheres are not necessarily limited to single white dwarfs \citep{zuckerman2014},
we have probed
stability limits in evolving circumstellar two-planet systems with non-evolving
secondaries. We found that for circular binaries, if the planets 
are sufficiently far from their parent star to avoid engulfment on the giant branch phase, 
then $a_{\rm B}(0)/a_2(0) \gtrsim 7$ ensures stability across all phases of 
stellar evolution. Instability during the white dwarf phase occurs for an increasing
spread of $a_{\rm B}(0)/a_2(0)$ values as $e_{\rm B}$ increases. Our results
(Fig. \ref{mainplot}) (i) provide useful benchmarks from which one could set up simulations
that generate instability along particular phases, (ii) display phase space portraits which
can be compared to observed systems, and (iii) represent a
starting point for explorations of more complex architectures with asteroids, moons, additional 
planets and/or significantly inclined bodies.

\section*{Acknowledgements}

We thank the referee, Cristobal Petrovich, for his insightful and probing feedback, which has improved the manuscript. DV and BTG have received funding from the European Research Council under the European Union's Seventh Framework Programme (FP/2007-2013)/ERC Grant Agreement n. 320964 (WDTracer). We would like to thank the High Performance Computing Resources team at New York University Abu Dhabi and especially Jorge Naranjo for helping us carry out our simulations.




\label{lastpage}
\end{document}